\documentclass[manuscript]{emulateapj}
\usepackage{natbib}
\usepackage{color}
\usepackage{amsmath}
\usepackage{array}
\newcolumntype{P}[1]{>{\centering\arraybackslash}p{#1}}

\usepackage[toc,page]{appendix}
\usepackage{afterpage}
\usepackage{float}
\usepackage{capt-of}
\usepackage{bm}
\usepackage[dvipsnames]{xcolor}
\usepackage{esint}
\usepackage{ mathrsfs }
\usepackage{lineno}

\definecolor{darkmagenta}{rgb}{0.5, 0, 0.5}

\shortauthors{Carr \& Scarlata}
 
\begin{document}

\title{A Semi-Analytical Line Transfer (SALT) Model III: Galactic Inflows}
\author{C. Carr\altaffilmark{1}, C. Scarlata\altaffilmark{1}}

 \altaffiltext{1}{Minnesota Institute for Astrophysics, School of
   Physics and Astronomy, University of Minnesota, 316 Church str
 SE, Minneapolis, MN 55455,USA}

\begin{abstract}
We present calculations of ultraviolet spectra resulting from the scattering of photons by gas in-falling onto an isotropically emitting source of radiation.  The model is based on an adaptation of the semi-analytical line transfer (SALT) code of \cite{Scarlata2015}, and designed to interpret the inverse P-Cygni profiles observed in the spectra of partially ionized galactic inflows.  In addition to presenting the model, we explore the parameter space of the inflowing SALT model and recreate various physically motivated scenarios including spherical inflows, inflows with covering fractions less than unity, and galactic fountains (i.e., galactic systems with both an inflowing and outflowing component). The resulting spectra from inflowing gas show spectral features that could be misinterpreted as ISM features in low resolution spectroscopy ($\sigma \approx 120 \ \rm{km } \ \rm{s}^{-1}$), suggesting that the total number of galactic systems with inflows is undercounted. Our models suggest that observations at medium resolution  ($R = 6000$ or $\sigma \approx 50 \ \rm{km } \ \rm{s}^{-1}$) that can be obtained with 8m-class telescopes will be able to resolve the characteristic inverse P Cygni profiles necessary to identify inflows.     
\end{abstract}

\section{Introduction}

Galactic scale flows (both inflowing and outflowing) are an essential ingredient in models of galaxy formation.  In particular, flows are responsible for regulating the baryon cycle and are necessary to reproduce observations on both galactic and cosmic scales.  A simple, but useful explanation is that flows work in equilibrium.  For example, supernovae or AGN driven feedback is necessary to expel matter from the interstellar medium (ISM) to enrich the circumgalactic medium (CGM, \citealt{Tumlinson2017}) and intergalactic medium (IGM, \citealt{Oppenheimer2006}), and in doing so, actively suppresses star formation; consequently, accretion is then necessary to refuel star formation to - for example - recreate the observed time evolution of the star formation rate (e.g., \citealt{Erb2008,Mannucci2010,Papovich2011}).  In this paradigm, knowing the physical properties of flows is essential for testing models of galaxy formation against observations.  Indeed, flow parameters are often introduced into hydrodynamical simulations in an ad hoc fashion to incorporate the physics of star formation which occurs on unresolved scales (e.g., \citealt{Springel2005,Vogelsberger2014}).  Furthermore, simulations which rely on self consistent models of star formation to derive flows (e.g., \citealt{Hopkins2014,Hopkins2018}) need to be checked as well to ensure their results agree with observations.  For a review on the status of galaxy formation models which emphasizes the importance of observationally constraining flows see \cite{Somerville2015}.

Observationally, flows are identified by the asymmetric absorption and emission line broadening observed in their spectra where the width reflects the maximum velocity of the flow and the depth reflects the optical depth of the relevant transition \citep{Prochaska2011,Scarlata2015,Carr2018}.  Over the past decade, absorption and emission line studies have made tremendous progress constraining the properties of flows across multiple wavelengths.  For example, local UV studies relying on the Hubble Space Telescope's Cosmic Origins Spectrograph (COS) have successfully constrained the cool and warm phases of the CGM from low and high ionization state metal lines  \citep{Stocke2013,Werk2014,Peeples2014,Heckman2017,Carr2021}, while sub-milimeter studies with the Herschel Space Observatory's PACS instrument have constrained the cold phase of the CGM via atomic and molecular lines \citep{Gonzalez-Alfonso2012,Falstad2015,Herrera-Camus2020}.  Ground based studies in the optical with Keck and milimeter with ALMA have placed constraints on flows at even higher redshifts \citep{Martin2012,Rubin2014,Gallerani2018}.  The hot phase of the CGM is the least well constrained due to the difficulty in observing the faint x-ray emission produced at such high temperatures ($\sim 10^7$ K); however, evidence for outflows has been observed in the x-ray emission of nearby galaxies \citep{Strickland1997,Laha2018}.  Together these studies reveal that outflows are ubiquitous, multiphase, and energetic reaching speeds on the order of hundreds to a few thousand kilometers per second.  In contrast, inflows are less energetic, typically reaching speeds less than 200 kilometers per second, and are observed far less frequently.  For example, in a study of $\sim 200$ galaxies from $0.4 < z < 1.4$, \cite{Martin2012} found evidence for inflows in only $3-6\%$ of their sample compared to the roughly $20\%$ detection rate for outflows in the absorption lines of Fe II (i.e., partially ionized gas $\sim10^4$K) under similar restrictions on the net observed velocity.  For a summary on the current state of observations of both inflows and outflows see \cite{Roberts-Borsani2019} and \cite{Veilleux2020}.

Given the important role accretion plays in models of galaxy formation, the overall lack of detections of inflows is surprising.  Indeed, hydrodynamical simulations predict that a significant amount of the gas feeding star formation at low redshifts is metal enriched \citep{Ford2014,Angles-Alcazar2017}.   For example, using the FIRE cosmological simulations, \cite{Angles-Alcazar2017} found that all galaxies in their simulations re-accrete $>50\%$ of the gas ejected in winds.  A possible explanation for the lack of detections is that galactic inflows occur along streams with low covering fractions, and hence, whether or not inflowing gas is detected will be highly sensitive to the viewing angle  \citep{Kimm2011}.  Furthermore, the spectral features of the more energetic outflows may mask that of inflows \citep{Martin2012,Rubin2014}.  For example, the symmetric emission component of an outflow may fill in the weaker redshifted absorption feature necessary to identify inflows in resonant lines.    

The observational approach to study the CGM (i.e., background vs down the barrel spectroscopy) matters in terms of what information can be recovered from flows and how efficiently.  In the case of background spectroscopy, where light from a distant quasar is used to probe the CGM of a single galaxy, the orientation of the flow cannot be determined from the absorption lines alone.  When this is the case, the galaxy's morphology and orientation can be used to predict the direction of the flow; however, these analyses suffer from model degeneracies and favor certain orientations for constraining flow properties (e.g., \citealt{Ho2020}).  Down-the-barrel spectroscopy, on the other hand, has the potential to unambiguously identify the orientation of a flow and has been used to study the spectra of outflows (e.g., \citealt{Rivera-Thorsen2015,Zhu2015,Chisholm2017a,Gazagnes2018,Carr2021}) and to estimate mass outflow rates (e.g., \citealt{Chisholm2016,Chisholm2017b,Xu2022}).  There is, however, very little modeling available to study the spectra of inflows in this area.  The lack of modeling has been recognized (see \citealt{Faucher2017}), and any progress made would be a step forward.            

In this paper, we present a modified version of the semi-analytical line transfer (SALT) model of \cite{Scarlata2015} to predict the spectra of galactic inflows obtained via down the barrel spectroscopy.  The model pertains only to the scattering of continuum radiation, and does not include emission from the inflows themselves.  As such, it cannot be used to interpret - for example - radio observations, but can be used to interpret the spectra of cool partially ionized gas where collisional excitation is expected to be low (e.g., \citealt{Rubin2012}).  We plan to consider an emission component in a future paper.

The rest of this paper is organized as follows.  In section 2, we adapt the SALT model of \cite{Scarlata2015} to account for galactic inflows.  In section 3, we explore the parameter space of the inflowing SALT model in various physically motivated scenarios including spherical inflows, inflows with covering fractions less than unity, and galactic fountains.  Our discussion and conclusions follow in section 4.   

\section{Modeling} 
\label{Modeling}

This paper is the third in a series of theoretical papers aimed at calculating the spectra of galactic flows.  In the first paper, \cite{Scarlata2015} presented a semi-analytical line transfer (SALT) model to predict the spectra of spherical galactic outflows.  Their work was later generalized by \cite{Carr2018} to include spectra of bi-conical outflows.  The aim of the current paper is to adapt SALT to predict the spectra of galactic inflows (both spherical and with covering fractions less than unity).  This model is intended to interpret the inverse P Cygni profiles observed in the UV lines associated with the low ionization metals (e.g., Mg II, Si II, etc.) thought to trace cool inflows during metal recycling (e.g., \citealt{Ford2014}).

\begin{figure}
  \centering
\includegraphics[scale=.4]{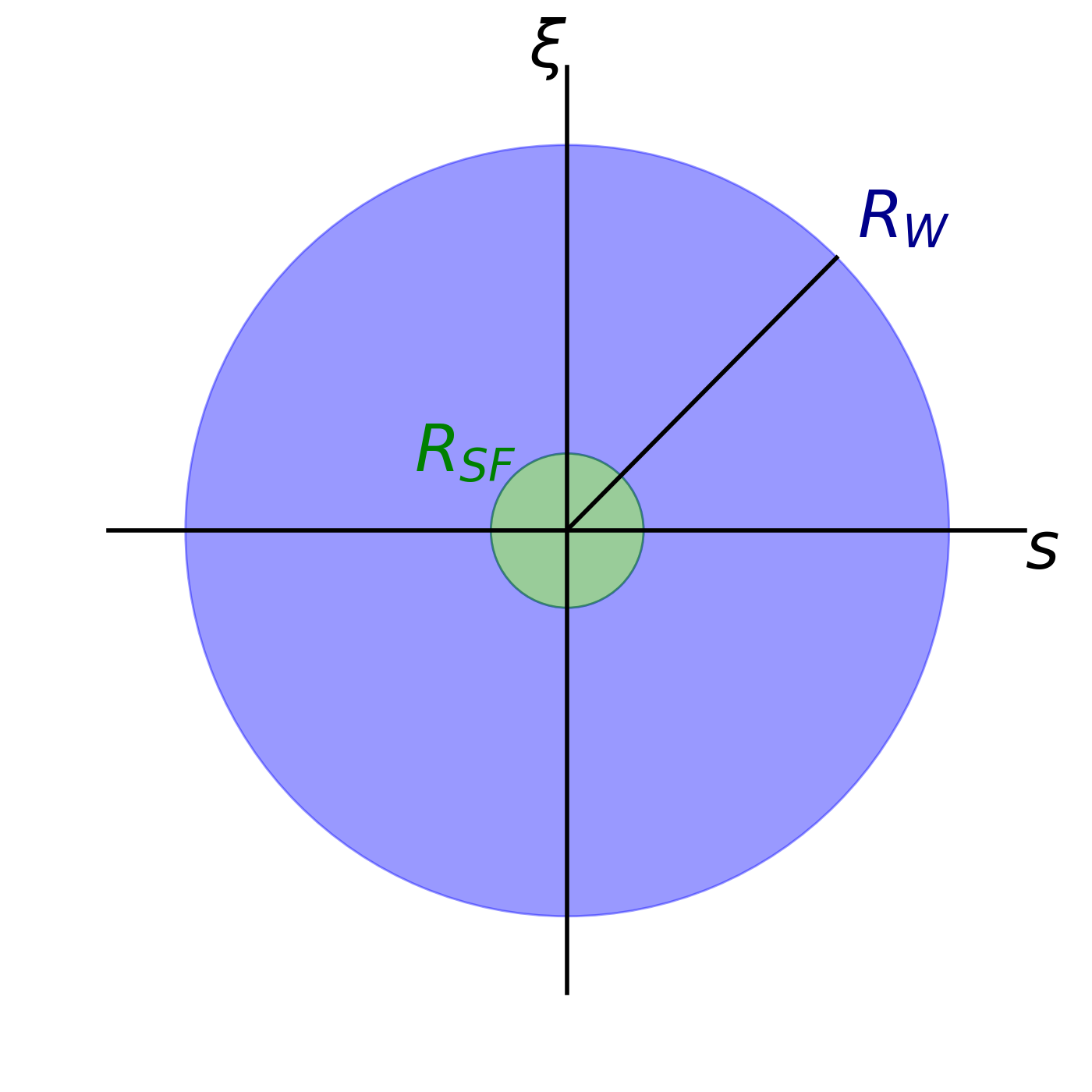}
 \caption{Envelope of material (shown in blue) of radius $R_W$ surrounding a galaxy (shown in green) of radius $R_{SF}$.  The $\xi$ and $s$-axes are written in normalized units $(r/R_{SF})$.  The envelope represents a flow characterized by a density and velocity field.}
   \label{f1}
\end{figure}

For consistency, we preserve what notation we can from previous works.  All models are to assume a spherical source of isotropically emitted radiation of radius, $R_{SF}$, surrounded by an envelope of flowing material which extends to a terminal radius, $R_W$.  A diagram has been provided in Figure~\ref{f1}.  The $\xi$-axis runs perpendicular to the line of sight and is measured using normalized units, i.e., $\xi=r/R_{SF}$.  The $s$-axis runs parallel with the line of sight and is measured using the same normalized units.    

For convenience, we have provided a complete list of the SALT parameters for the outflowing and inflowing SALT model along with their definitions in Table~\ref{tab:pars}.  More detailed definitions can be found in \cite{Carr2018,Carr2021}. 

\subsection{SALT Model for Galactic Inflows}

We begin by assuming the velocity field of the in-falling gas follows a power law of the form,
\begin{equation}
\begin{aligned}
v &= v_{\infty} - v_0\left(\frac{r}{R_{SF}}\right)^{\gamma}&&\text{for}\ r < R_{W} \\[1em]
v &= 0 &&\text{for} \ r \geq R_{W}, \\
\end{aligned}
\end{equation}     
where $v = v_{\infty}-v_0$ represents the velocity at $R_{SF}$ and $\gamma$ takes positive values.  To ensure an inflow, we set the orientation of the velocity field to be positive along the line of sight when positioned between the observer and the source (i.e., opposite the orientation of the outflowing case).  This velocity field assumes that accretion begins from rest at some finite distance, $R_W$, away from the galaxy.  Such a situation is suitable for metal enriched gas that has been thrown out of the galaxy, comes to a halt, and falls back in.  Aside from this insight, the exact expression for the velocity field was chosen for analytical simplicity, and we direct the reader to \citealt{Fielding2022} for a discussion on the physics influencing the velocity fields of flows.  We assume an arbitrary density field, $n(r) = n_0(r/R_{SF})^{-\delta}$, for the in-falling medium.  The Sobolev approximation is still valid in this context \citep{Lamers1999}, and we assume photons can only interact locally with the inflow at a single point.  Staying consistent with prior works, we will use the normalized units: $y = v/v_0$ and $x = v \cos{\theta}/v_0$ where $v\cos{\theta}$ is the projection of $v$ onto the line of sight.  

\begin{figure}[]
  \centering
\includegraphics[scale=0.48]{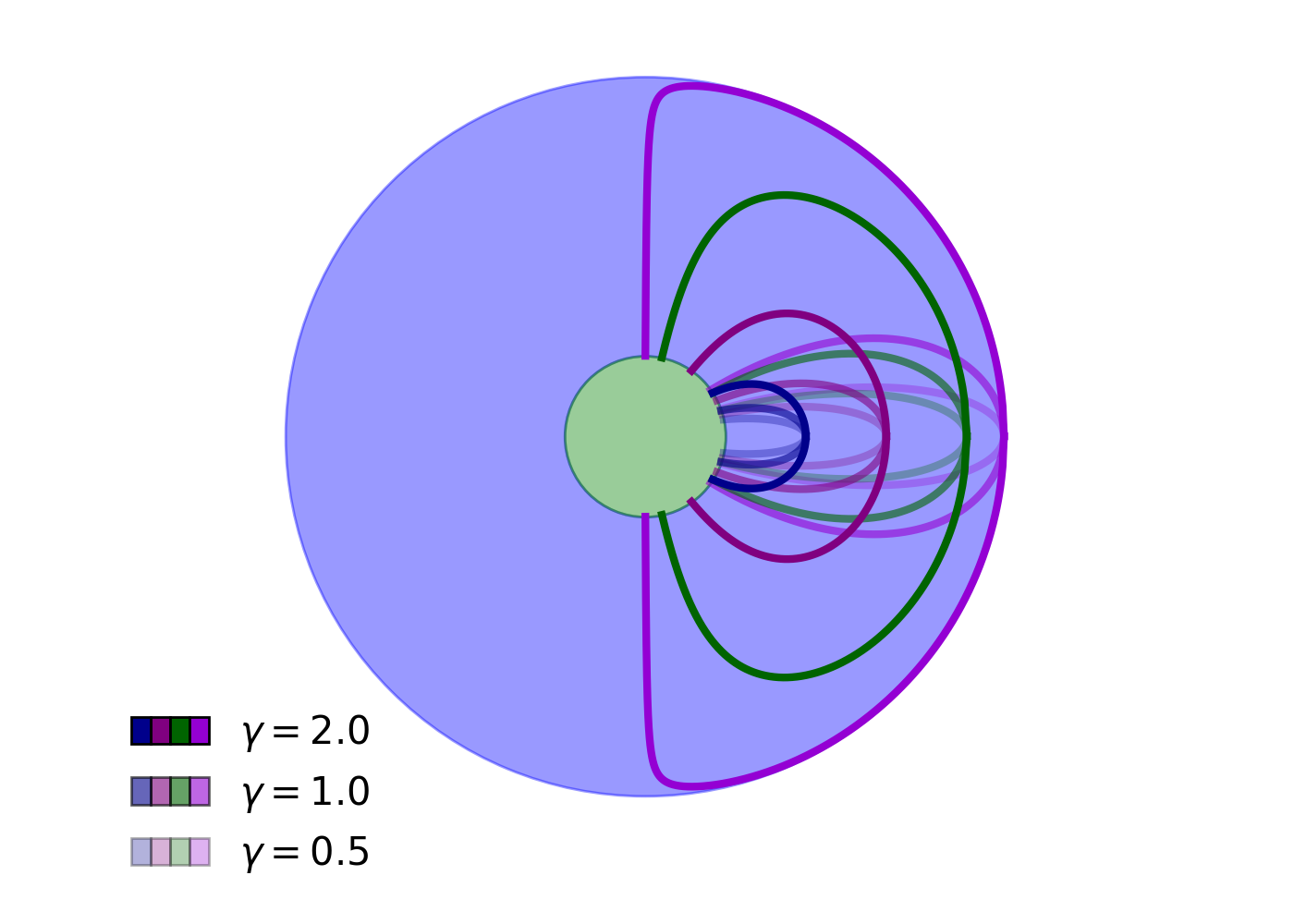}
 \caption{Various curves of constant observed velocity for inflowing velocity fields. The curves drawn with the darkest, middle, and lightest shades correspond to velocity fields with $\gamma = 2.0,1.0$, and $0.5$, respectively.  The different colors represent different observed velocities.}  
   \label{f2}
\end{figure}

Following the derivation of the outflowing SALT model in Carr et al. (2022, in prep), we seek to identify the surfaces of constant observed velocity, $\Omega_x$, in the inflow; however, due to radial symmetry, it will suffice to construct $\Gamma_x$ (i.e., the intersection of $\Omega_x$ with the $s\xi$-plane).  Following the derivation of \cite{Carr2018}, we compute  

\begin{eqnarray}
\Gamma_x(y) &&\resizebox{0.38\textwidth}{!}{$= \left([y_{\infty}-y]^{1/\gamma}\left(\frac{x}{y}\right),\left[(y_{\infty}-y)^{2/\gamma}-(y_{\infty}-y)^{2/\gamma}\left(\frac{x}{y}\right)^2\right]^{1/2}\right)$}\\
&&= (S,\Xi),
\end{eqnarray} 

\noindent where $S(y)$ and $\Xi(y)$ refer to the parameterizations of the $s$ and $\xi$ coordinates of $\Gamma_x$, respectively.  Several different examples of $\Gamma_x$ are shown in Figure~\ref{f2} for different values of $x$ and $\gamma$.  An interesting difference between these curves and those of an outflowing velocity field\footnote{Note this claim is making the additional assumption that the outflow is accelerating.  For decelerating velocity fields, $\Gamma_x$ displays the same behavior as inflowing velocity fields.} (see \citealt{Carr2018}), is that all curves for an inflowing velocity field trace back to the source (i.e., towards regions of high density).  As we will demonstrate, this is one of the reasons why the line profile for an inflow gives the appearance of an 'inverse' P Cygni profile.  

\begin{figure}[]
  \centering
\includegraphics[scale=0.34]{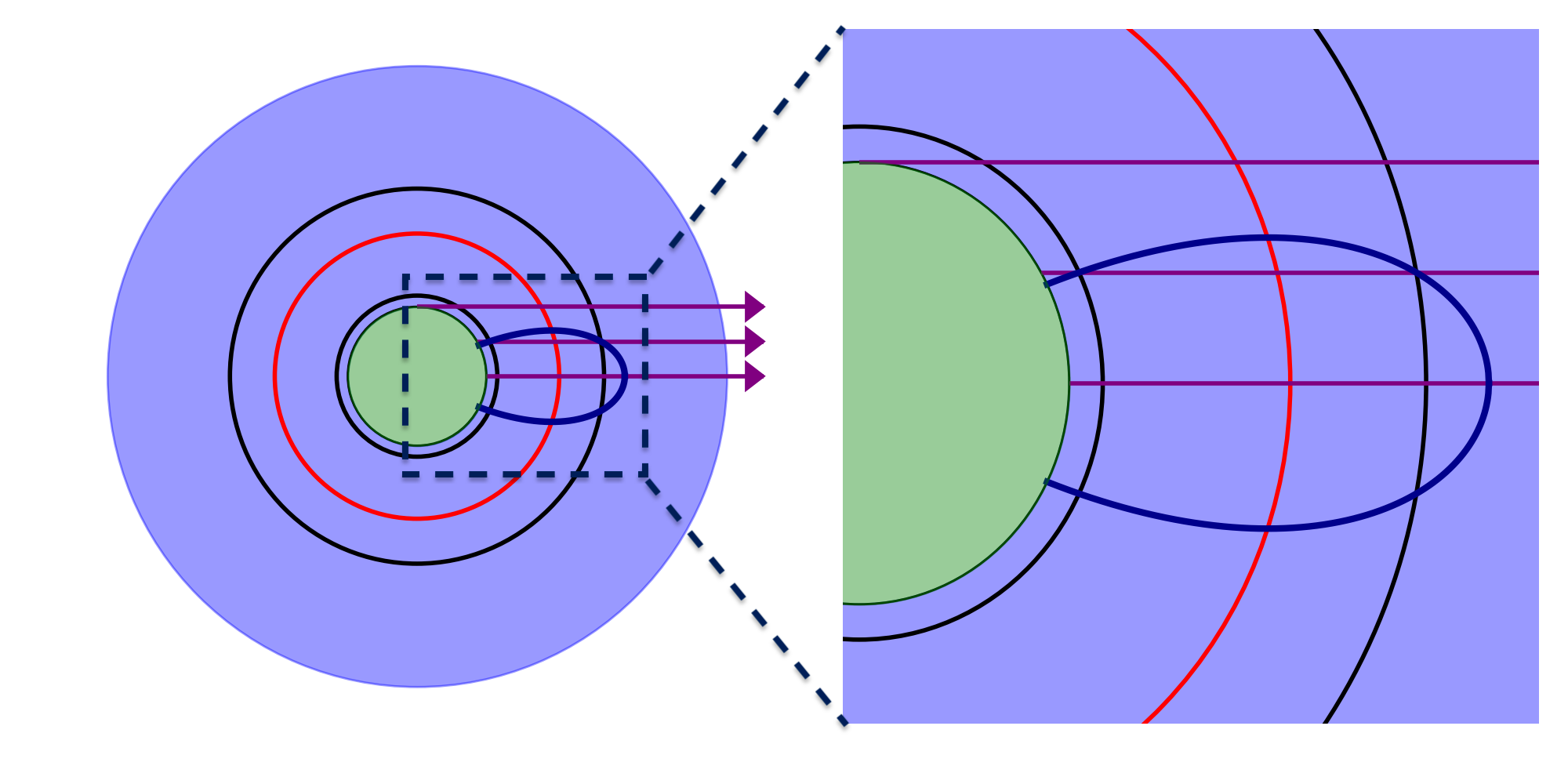}
 \caption{\emph{Left} Crosssectional view of a spherical inflow.  Continuum rays are drawn in purple and are shown passing through a surface of constant observed velocity whose curve of constant observed velocity, $\Gamma_x$, is shown in blue.  Note that the rays can either intersect $\Gamma_x$ once, twice, or zero times.  For the continuum ray which intersects $\Gamma_x$ twice, the shells of intrinsic velocities $y_{h_1}$ and $y_{h_2}$ which pass through the intersections are shown in black.  A shell of intrinsic velocity, $y_{max}$, intersecting the maximum of $\Gamma_x$, is shown in red.  \emph{Right} A close up view of the intersections.  The maximum of $\Gamma_x$ as well as the points of intersection with the continuum rays provide enough information to compute the absorption profile as described in the text.   }
   \label{f3}
\end{figure}

Observe that rays emitted parallel to the line of sight can intersect a surface of constant observed velocity once, twice, or zero times.  An example of a surface for which all three scenarios are possible is shown in Figure~\ref{f3}.  This issue does not occur in an outflowing velocity field and requires special attention.  Consider a continuum ray emitted at a height, $h$, above the $s$-axis.  To determine if this ray will pass through the surface of constant observed velocity, $\Omega_x$, we must first find the maximum height, $\Xi(y_{max})$, reached by $\Gamma_x$ in the $s\xi$-plane.  We have   

\begin{eqnarray}
\resizebox{0.41\textwidth}{!}{$\frac{d\Xi}{dy} = \frac{\gamma x^2y^{-3}[y_{\infty}-y]^{2/\gamma}+(x^2y^{-2}-1)[y_{\infty}-y]^{(2-\gamma)/\gamma}}{-\Xi([y_{\infty}-y]^{(1-\gamma)/\gamma}xy^{-1}+\gamma[y_{\infty}-y]^{1/\gamma}xy^{-2})}$,}
\end{eqnarray}
and after minimizing we deduce that
\begin{eqnarray}
0 &=& y^3+(\gamma-1)x^2y-\gamma y_{\infty}x^2,
\end{eqnarray}
which has the real valued solution
\begin{eqnarray}
\resizebox{0.41\textwidth}{!}{$y_{\rm{max}} = \left(-\frac{q}{2}+\left(\frac{q^2}{4}+\frac{p^3}{27}\right)^{1/2}\right)^{1/3}+\left(-\frac{q}{2}-\left(\frac{q^2}{4}+\frac{p^3}{27}\right)^{1/2}\right)^{1/3}$,}
\end{eqnarray}
where $p = (\gamma-1)x^2$ and $q = -\gamma x^2y_{\infty}$.  Thus, if $\Xi(y_{max}) < h$, then the continuum ray will miss $\Omega_x$.  A shell with intrinsic velocity $y_{max}$ which intersects $\Omega_x$ at its maximum is shown in red in Figure~\ref{f3}.    

If $\Xi(y_{\infty}-1) < h < \Xi(y_{max})$, then the continuum ray will intersect $\Omega_x$ twice.  Setting $\Xi = h$, we deduce the following relation,
\begin{eqnarray}
\frac{h^2}{(y_{\infty}-y)^{2/\gamma}}=1-\left(\frac{x}{y}\right)^2,
\label{intersections}
\end{eqnarray}
which has solutions $y_{h_1}$ and $y_{h_2}$ corresponding to the intrinsic velocities of the shells intersecting $\Gamma_x$ at height $h$.  Shells with intrinsic velocities $y_{h_1}$ and $y_{h_2}$ for a fixed value of $h$ are shown in black in Figure~\ref{f3}.  Finally, if $h < \Xi(y_{\infty}-1)$ the continuum ray will intersect $\Omega_x$ only once.   

Putting it all together, we compute the absorption spectrum as follows.  We define the integrand, $\rm{ABS}$, as  

\begin{eqnarray}
\rm{ABS} &=& 2 \pi h (1-e^{-\bar{\tau}_S})|dh/dy|/\pi \\
&=&\resizebox{0.35\textwidth}{!}{$\frac{2}{\gamma}\left[\frac{x^2}{y^2}(y_{\infty}-y)^{\frac{2-\gamma}{\gamma}}+\frac{\gamma x^2}{y^3}(y_{\infty}-y)^{\frac{2}{\gamma}}-(y_{\infty}-y)^{\frac{2-\gamma}{\gamma}}\right]$}\nonumber\\
&\times& (1-e^{-\bar{\tau}_S(y)}),
\label{SF_abs1}
\end{eqnarray}
where $\bar{\tau}_S(y)$ is defined over the following cases.

\begin{figure*}[]
  \centering
\includegraphics[scale=0.58]{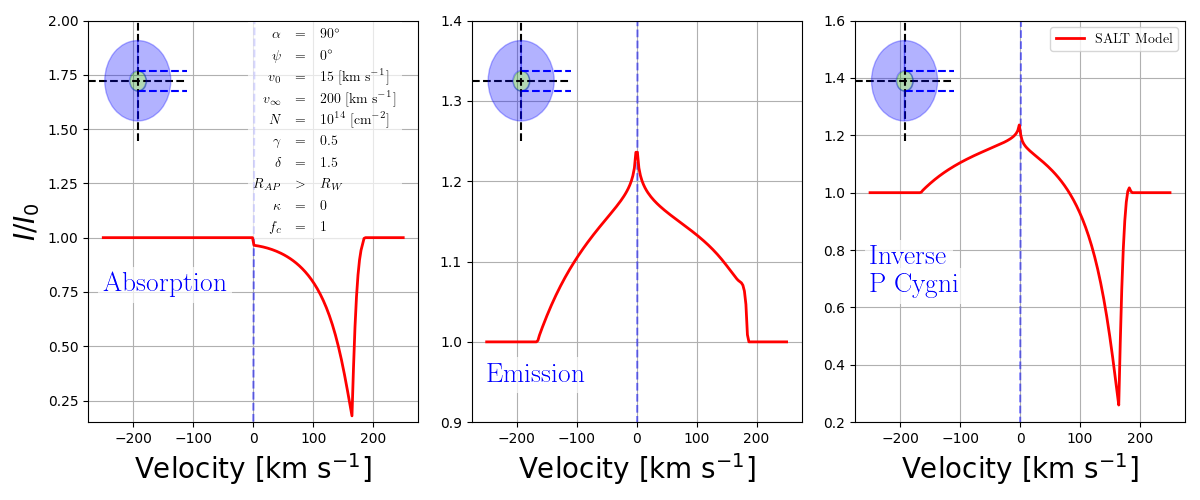}
 \caption{\textbf{\emph{Left}} The absorption profile, \textbf{\emph{Middle}} emission profile, and \textbf{\emph{Right}} Inverse P Cygni profile for a spherical inflow.  The asymmetry in the emission profile reflects the occultation effect where photons emitted from behind the source are blocked by the source from the observer's field of view.  The parameters for this inflow are listed on the left most panel.}
   \label{f4}
\end{figure*}

\begin{gather} 
\bar{\tau}_S \equiv 
\begin{cases}
\tau_S(y_{h_1})+\tau_S(y_{h_2})&\ \rm{if} \  \Xi(y_{\infty}-1)\leq \Xi(y)< \Xi(y_{\rm{max}}) \\[1em]
\tau_S(y) & \ \rm{otherwise}\\
\end{cases}
\label{SF_abs2}
\end{gather}
and $\tau_S$ is the Sobolev optical depth \citep{Lamers1999} defined as
\begin{eqnarray}
\tau_S &=& \frac{\pi e^2}{mc} f_{lu}\lambda_{lu} n_l(r)\left[1-\frac{n_ug_l}{n_lg_u}\right] \frac{r/v}{1+\sigma {\cos^2{\phi}}},
\label{Castor}
\end{eqnarray}
where $f_{ul}$ and $\lambda_{ul}$ are the oscillator strength and wavelength, respectively, for the $ul$ transition, $\sigma = \frac{d \ln(v)}{d\ln(r)} - 1$, and $\phi$ is the angle between the velocity and incoming photon ray \citep{Castor1970}.  All other quantities take their usual definition.  Writing Equation~\ref{Castor} in terms relevant to the SALT model, and by neglecting stimulated emission (i.e., $\left[1-\frac{n_ug_l}{n_lg_u}\right] =1$), we get
\begin{eqnarray}
\tau_S(y) = \tau_0 \frac{(y_{\infty}-y)^{(1-\delta)/\gamma}}{y+[\gamma(y_{\infty}-y)-y](x/y)^2]},
\label{inflow_tau}
\end{eqnarray}
where
\begin{eqnarray}
\tau_0 = \frac{\pi e^2}{mc} f_{lu}\lambda_{lu} n_0 \frac{R_{SF}}{v_0}.
\end{eqnarray}
Note that we are assuming that the population of the excited state is negligible ($n_u\sim0$), and are therefore excluding the possibility of emission from the inflow and are focusing exclusively on continuum scattering.  We will consider an emission component in a future paper and extend the range of our model to include radio observations and the interpretation of the inverse P Cygni profiles frequently observed in the spectra of cool atomic and molecular inflows (e.g., \citealt{Gonzalez-Alfonso2012,Falstad2015,Herrera-Camus2020}).

Finally, if $\Xi(y_{max})\leq1$, then the normalized absorption profile becomes: 

\begin{eqnarray}
\resizebox{0.4\textwidth}{!}{$
\frac{I_{\rm{abs,red}}}{I_0} = \frac{F(x)}{F_c(x)} - \frac{F(x)}{F_c(x)}\int_{x}^{\rm{min}(y_{max},y_{\infty}-1)}\rm{ABS}(y) \ dy,$}
\end{eqnarray}
where $F(x)/F_c(x)$ is the flux at $x$ normalized by the continuum flux.  If $\Xi(y_{max})>1$, then the normalized absorption profile becomes 
\begin{eqnarray}
\frac{I_{\rm{abs,red}}}{I_0}= \frac{F(x)}{F_c(x)}  - \frac{F(x)}{F_c(x)} \int_{x}^{y_h}\rm{ABS}(y) \ dy, 
\end{eqnarray}
where $y_h = \rm{min}(y_{h_2}(1),y_{h_1}(1))$, and $y_{h_2}(1)$ and $y_{h_1}(1)$ are the solutions of equation~\ref{intersections} for $h=1$.

We can compute the emission component of the inflowing SALT model using the same general procedure used in the outflowing case (see Carr et al. 2022, in prep).  To this end, we think of $\Gamma_x(y)$ as a homotopy (i.e., a continuous map between surfaces of constant observed velocity), and compute the amount of energy absorbed by each surface that lies within the shell as we move in observed velocity space from $x = y\cos{\Theta_C}$ to $x = y$ where $\Theta_C = \arcsin{\left([y_{\infty}-y]^{-1/\gamma}\right)}$.  Similar to the absorption case, we must be careful to identify which rays intersect $\Omega_x(y)$ before reaching the shell.  With this in mind, we compute the total energy absorbed by a shell at velocity $y$ in the inflow as
\begin{eqnarray}
&&\hspace*{-1cm}L_{\rm{shell}} = \nonumber \\
&&\hspace*{-1cm}\resizebox{0.43\textwidth}{!}{$ \int_{y\cos{\Theta_C}}^{y}\frac{2L(x)}{\gamma}\left[\frac{x^2}{y^2}(y_{\infty}-y)^{\frac{2-\gamma}{\gamma}}+\frac{\gamma x^2}{y^3}(y_{\infty}-y)^{\frac{2}{\gamma}}-(y_{\infty}-y)^{\frac{2-\gamma}{\gamma}}\right] \rm{SF} dx,$}
\end{eqnarray}
where $L(x)$ is the total energy emitted by the shell at resonance with material moving at observed velocity, $x$.  We refer to SF as the Sobolev factor and define it for the following cases.  If $y \geq y_{max}$, then
\begin{eqnarray}
\rm{SF} \equiv -[1-e^{-\tau_S(y)}];
\label{neg_change}
\end{eqnarray}
otherwise,

\begin{gather} 
\resizebox{0.48\textwidth}{!}{$
\rm{SF} \equiv 
\begin{cases}
e^{-\tau_S(y_{h_2})}[1-e^{-\tau_S(y_{h_1})}]& \ \rm{if} \  \Xi(y_{\infty}-1)\leq \Xi(y)< \Xi(y_{\rm{max}}) \\[1em]
1-e^{-\tau_S(y)} & \ \rm{otherwise,}\\
\end{cases}
$}
\label{SF_em}
\end{gather}
where $\tau_S$ is given by Equation~\ref{inflow_tau}.  The leading negative sign in Equation~\ref{neg_change} accounts for the sign change in $dh$ when $y>y_{max}$.  Now that we know the total energy absorbed by the shell, the reemission of energy in terms of the observed velocities can be computed exactly as in the outflowing case, albeit for a change in orientation (see Carr et al. 2022, in prep).  We compute the normalized red emission profile as 
\begin{eqnarray}
&&\hspace*{-.9cm}I_{\rm{em,red}}/I_0 = \nonumber \\
&&\hspace*{-.9cm}\resizebox{0.43\textwidth}{!}{$\int_{x}^{y_{\infty}-1} \frac{F(x^{\prime})}{F(x)}\frac{dy}{2y} \int_{y\cos{\Theta_C}}^{y}\frac{2}{\gamma}\left[\frac{{x^{\prime}}^2}{y^2}(y_{\infty}-y)^{\frac{2-\gamma}{\gamma}}+\frac{\gamma {x^{\prime}}^2}{y^3}(y_{\infty}-y)^{\frac{2}{\gamma}}-(y_{\infty}-y)^{\frac{2-\gamma}{\gamma}}\right] \rm{SF} dx^{\prime},$}
\end{eqnarray}
where $SF$ is defined for the cases given above.  Similar to the outflowing case, we account for the occultation effect on the blue emission profile by removing all photons reemitted from behind the source at heights $h = \Xi < 1$.  To this end, we define the scale factor  $\Theta_{\rm{blue}} \equiv \Theta(\Xi-1)$ such that 
\begin{gather} 
\Theta \equiv 
\begin{cases}
1 & \rm{if \ }  \ \Xi > 1 \\[1em]
0 & \rm otherwise.\\
\end{cases}
\end{gather}
The normalized blue emission profile becomes   
\begin{eqnarray}
&&\hspace*{-.9cm}I_{\rm{em,blue}}/I_0 = \nonumber \\
&&\hspace*{-.9cm}\resizebox{0.43\textwidth}{!}{$\int_{x}^{y_{\infty}-1} \Theta_{\rm{blue}}\frac{F(x^{\prime})}{F(x)}\frac{dy}{2y} \int_{y\cos{\Theta_C}}^{y}\frac{2}{\gamma}\left[\frac{{x^{\prime}}^2}{y^2}(y_{\infty}-y)^{\frac{2-\gamma}{\gamma}}+\frac{\gamma {x^{\prime}}^2}{y^3}(y_{\infty}-y)^{\frac{2}{\gamma}}-(y_{\infty}-y)^{\frac{2-\gamma}{\gamma}}\right] \rm{SF} dx^{\prime}.$}
\end{eqnarray}  

The technique developed by \cite{Carr2018} to account for the effects of a biconical outflow geometry still holds and can be applied to the line profile in the obvious way; however, rays which can interact with a surface of constant observed velocity more than once must be handled carefully since the geometry of the flow may be different at each interaction.  When this is the case, we can approximate the expression for $SF$ in Equation~\ref{SF_em} as 
\begin{eqnarray}
f_g(y_{h_1})[I(y_{h_2})-I(y_{h_2})e^{-\tau_S(y_{h_1})}],
\label{new_g}
\end{eqnarray}
where 
\begin{eqnarray}
I = 1-f_g(y_{h_2})[1-e^{-\tau_S(y_{h_2})}],
\end{eqnarray}
and $f_g$ is the geometric factor as defined by \cite{Carr2018}.  Likewise, the Sobolev factor defined in Equations~\ref{SF_abs1} \& \ref{SF_abs2} will also need to change to the expression in Equation~\ref{new_g}, but with the addition of the term, $f_g(y_{h_2})[1-e^{-\tau_S(y_{h_2})}]$.  The techniques developed by \cite{Carr2018,Carr2021} to account for a dusty CGM, a limiting observing aperture, and holes in the outflow still hold and can be applied in the obvious way.  Lastly, the multiple scattering procedure of \cite{Scarlata2015} used to account for resonant and fluorescent reemission still applies. 

The absorption, emission, and full line profile or inverse P Cygni profile for a spherical inflow are shown in Figure~\ref{f4}.  The most striking difference when compared to the traditional line profiles of outflows is that absorption now occurs to the right of systemic velocity or at positive observed velocities using our orientation.  The emission profile is now more sharply peaked with a wider base compared to that of an outflow.  This is because, in our model, the highest density material near the source is moving at the highest velocities - opposite of the outflowing case - thus the majority of energy is now spread over a larger observed velocity range; however, absorption (and reemission) can still occur in shells with low intrinsic velocities all the way to zero intrinsic velocity (outflows are cut off at the launch velocity) and accumulate emission into a sharp peak.     

\section{Results}  
In this section, we explore the parameter space of the inflowing SALT model by generating spectra for a variety of distinct physically motivated scenarios for the emergence of inflows in galactic systems.  In other words, we explore the range of applicability for detecting galactic inflows using down-the-barrel spectroscopy - an area where modeling has been especially lacking (see \citealt{Faucher2017}).  We consider the case of pure inflows - both spherical and with partial covering fractions - as well as galactic fountains.  In our fiducial model, we assume $v_{\infty} = 200 \ \rm{km} \ {s}^{-1}$ and a column density ranging from $14 \leq \log{(N \ [\rm{cm^{-2}}])} \leq 15$.  These values were chosen to be in agreement with \cite{Roberts-Borsani2019} and  \cite{Martin2012}.   For the remaining parameters, we assume $\gamma = 0.5$, $\delta = 1.5$, $\kappa = 0.0$ (i.e., dust free), $v_0 = 15 \ \rm{km} \ {s}^{-1}$, and $R_{AP} > R_W$ (i.e., flow is fully captured by the observing aperture) while $\alpha$, $\psi$, and $f_c$ vary.  For the chosen velocity field, a value of $\gamma \leq 1$ ensures that the acceleration ($\sim vdv/dr$) of the inflow increases monotonically with decreasing radius, and a value of $\delta < 2$ ensures that the mass inflow rate ($\sim r^2nv$) will start to decrease with decreasing radius some finite distance from the source.  The latter can be expected since material is likely to be stripped from the flow during infall, and the ionizing flux should increase as the material moves closer to the galaxy thereby increasing the ionization state.

\subsection{Pure Inflows} 
By far the most convincing evidence for inflows in absorption line studies is the direct detection of inverse P Cygni profiles (e.g., \citealt{Rupke2021}).  These line profiles are characterized by having an overall net blueshift in emission and an overall net redshift in absorption.  The latter of which is a stark quality unique to the spectra of inflowing gas; however, to declare the detection of an inflow, the redshifted absorption must reach observed velocities exceeding the limitations of the spectral resolution of the study.  Furthermore, the absorption must be asymmetric with respect to line center (e.g., \citealt{Martin2012}) to distinguish the inflow from possible ISM absorption or thermal/turbulent line broadening (Carr et al. 2022, in prep).  Here we investigate which parameter ranges favor detections of inflows in absorption for cases of pure spherical inflows and inflows with covering fractions less than unity.    

\begin{figure*}[]
  \centering
\includegraphics[scale=0.5]{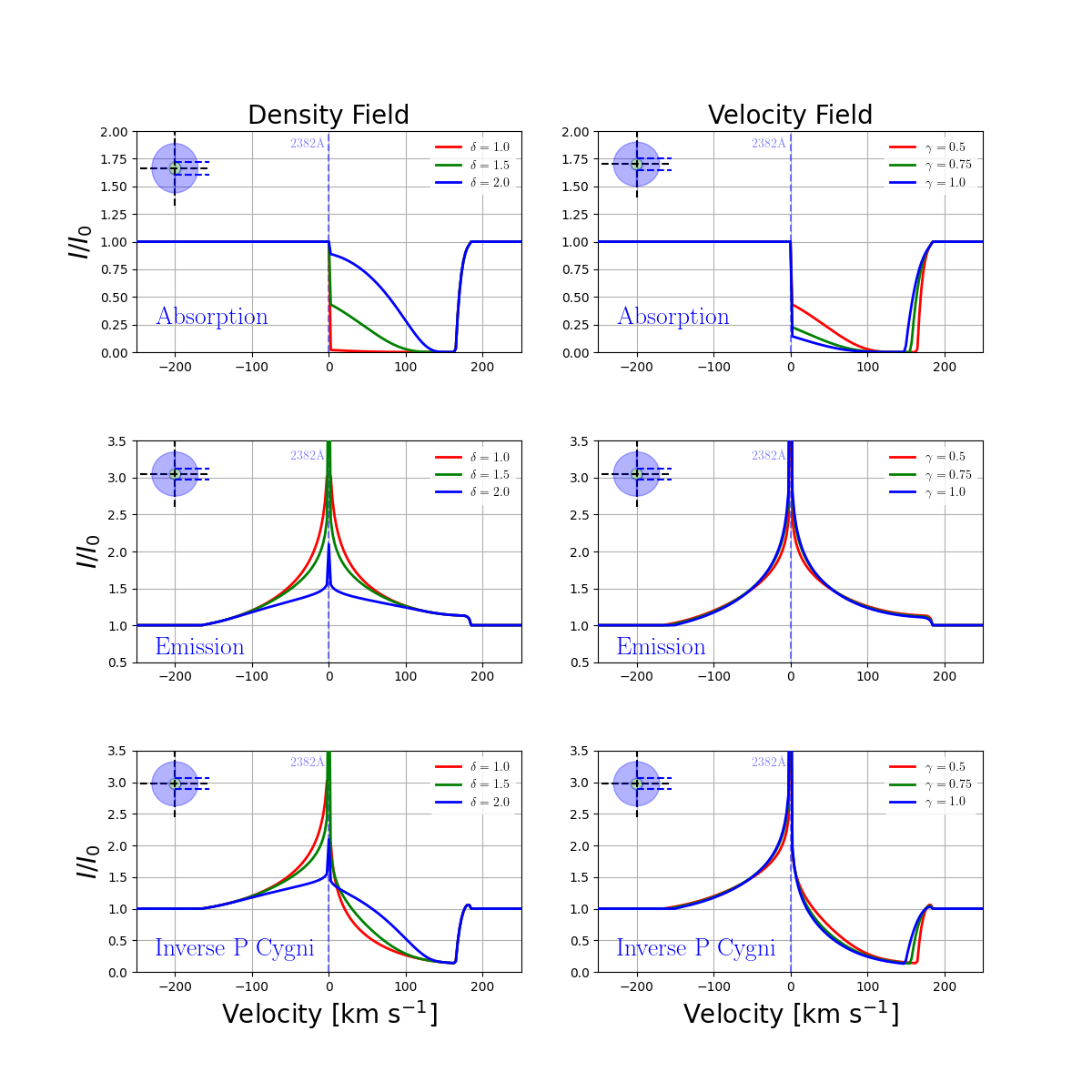}
 \caption{\textbf{\emph{Left Column}} The absorption profile (top), emission profile (middle), and  Inverse P Cygni profile (bottom) of a spherical inflow for various values of $\delta$ with $\gamma = 0.5$.  \textbf{\emph{Right Column}} The same as the left column except $\gamma$ varies and $\delta = 1.5$.  The remaining parameters for these profiles are $\alpha = 90^{\circ}$, $\psi = 0^{\circ}$, $\log{(N \ [\rm{cm}^{-2}])}=15$, $v_0 = 15 \ \rm{km \ s^{-1}}$, $v_w = 200 \ \rm{km \ s^{-1}}$, $R_{AP} > R_W$, $\kappa = 0$, and $f_c = 1.0$.  }
   \label{f5}
\end{figure*}

\begin{figure*}[]
  \centering
\includegraphics[scale=0.55]{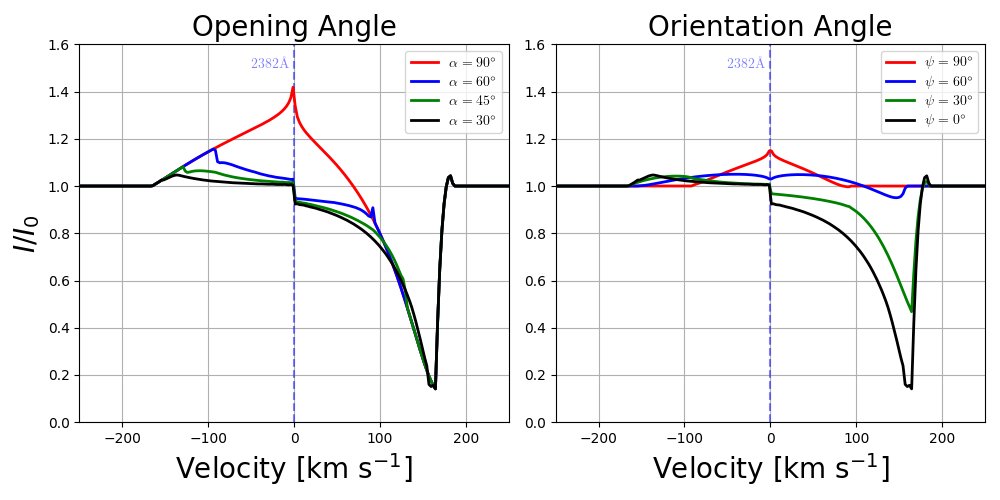}
 \caption{\textbf{ \emph{Left Panel}} Line profiles for biconical inflows of various opening angles oriented along the line of sight (i.e., $\psi = 0^{\circ}$).  \textbf{\emph{Right Panel}} Line profiles for biconical inflows with opening angle, $\alpha = 30^{\circ}$, at oriented at various angles with respect to the line of sight.  The remaining parameters for these profiles are $\gamma = 0.5$, $\delta = 1.5$, $\log{(N \ [\rm{cm}^{-2}])}=15$, $v_0 = 15 \ \rm{km \ s^{-1}}$, $v_w = 200 \ \rm{km \ s^{-1}}$, $R_{AP} > R_W$, $\kappa = 0$, and $f_c = 1.0$. }
   \label{f6}
\end{figure*} 

We examine the effects of varying the power laws of both the density field (left column) and velocity field (right column) on the line profile of a spherical inflow in Figure~\ref{f5}.  We have chosen to model the Fe II $2382$\AA \ resonant transition which was used in the study by \cite{Martin2012} to identify inflows and outflows in galaxies at redshifts $0.4<z<1.4$ as an example.  All atomic data for the lines used in this paper is provided in Table~\ref{tab:atomicdata}.  In each column, we show how the absorption (top panel), emission (middle panel), and full inverse P Cygni profile (bottom panel) vary as functions of $\delta$ (left column) and $\gamma$ (right column) for fixed column density, $\log{(N \ [\rm{cm}^{-2}])} = 15$.  In regards to density, the amount of absorption at low observed velocities increases with decreasing $\delta$.  This is because for steep density fields, most absorption is happening near the source at high observed velocities.  When the density field becomes  shallower, however, more absorption can occur at larger radii or at lower observed velocities.  The spectrum varies significantly less over our chosen range for $\gamma$, but absorption does appear to increases at lower observed velocities for increasing $\gamma$.  This occurs because the size of the inflow (i.e., $R_W = R_{SF}(v_{\infty}/v_0)^{1/\gamma}$) increases with decreasing $\gamma$.  Therefore, the bulk of material at low velocity is moving further away from the galaxy at lower density.  We can conclude that steeper density and shallower velocity fields yield better conditions for identifying inflows.  Under these conditions, most absorption will occur near the maximum observed velocity of the flow creating a more asymmetric (with regards to systemic velocity) absorption profile.

In Figure~\ref{f6}, we break the assumption of spherical symmetry and consider biconical inflows.  These are the same geometrical models used by \cite{Carr2018}, but are now inflowing.  We mention that this is only an example, and note that the SALT model can just as easily be used in the case of a single cone geometry.  In the left panel of Figure~\ref{f6}, we vary the opening angle, $\alpha$, of an inflow oriented along the line of sight.  As is the case with outflows (see \citealt{Carr2018}), there is less reemission for smaller opening angles.  This includes reducing the amount of red emission infilling\footnote{This is an analogous effect to the blue emission infilling which is typically associated with outflows.}, which enhances the increase of absorption for smaller opening angles.  The opposite effect occurs when increasing the orientation angle, $\psi$, which is shown for an inflow with an opening angle $\alpha = 30^{\circ}$ in the right panel of Figure~\ref{f6}.  As $\psi$ increases, the inflowing gas is moved away from the line of sight, which results in less absorption since there is less material between the observer and the source.  Coincidently, more and more emission appears near systemic or zero observed velocity, reflecting the position of the inflow with respect to the observer.  The importance of this example is to demonstrate that when spherical symmetry is broken either the absorption or emission equivalent width can dominate based on the orientation of the inflow with respect to the observer.  In regards to detecting inflows, a red shifted absorption profile is still a dead give away for inflowing gas; however, a pure emission profile could be mistaken for outflowing gas.  The only difference between emission profiles that arise from outflows and inflows, is that, in the case of inflows, they are much more narrow peaked with wider bases\footnote{Emission profiles with wider bases and narrow peaks are also characteristic of outflows with decelerating velocity fields.}.  In this instance, fitting these lines with a single Gaussian - as is typically done in more traditional modeling techniques - may be difficult.  More than likely fitting these emission lines would require two Gaussians (one for the narrow component near systemic velocity and the second for the wider base) and be misinterpreted as representing two kinematically distinct components.  Thus the necessity to fit emission profiles with more than one Gaussian could be evidence for inflowing gas.

\subsection{Outflow and Inflows: Galactic Fountains}

\begin{figure*}[]
 \centering
\includegraphics[scale=0.8]{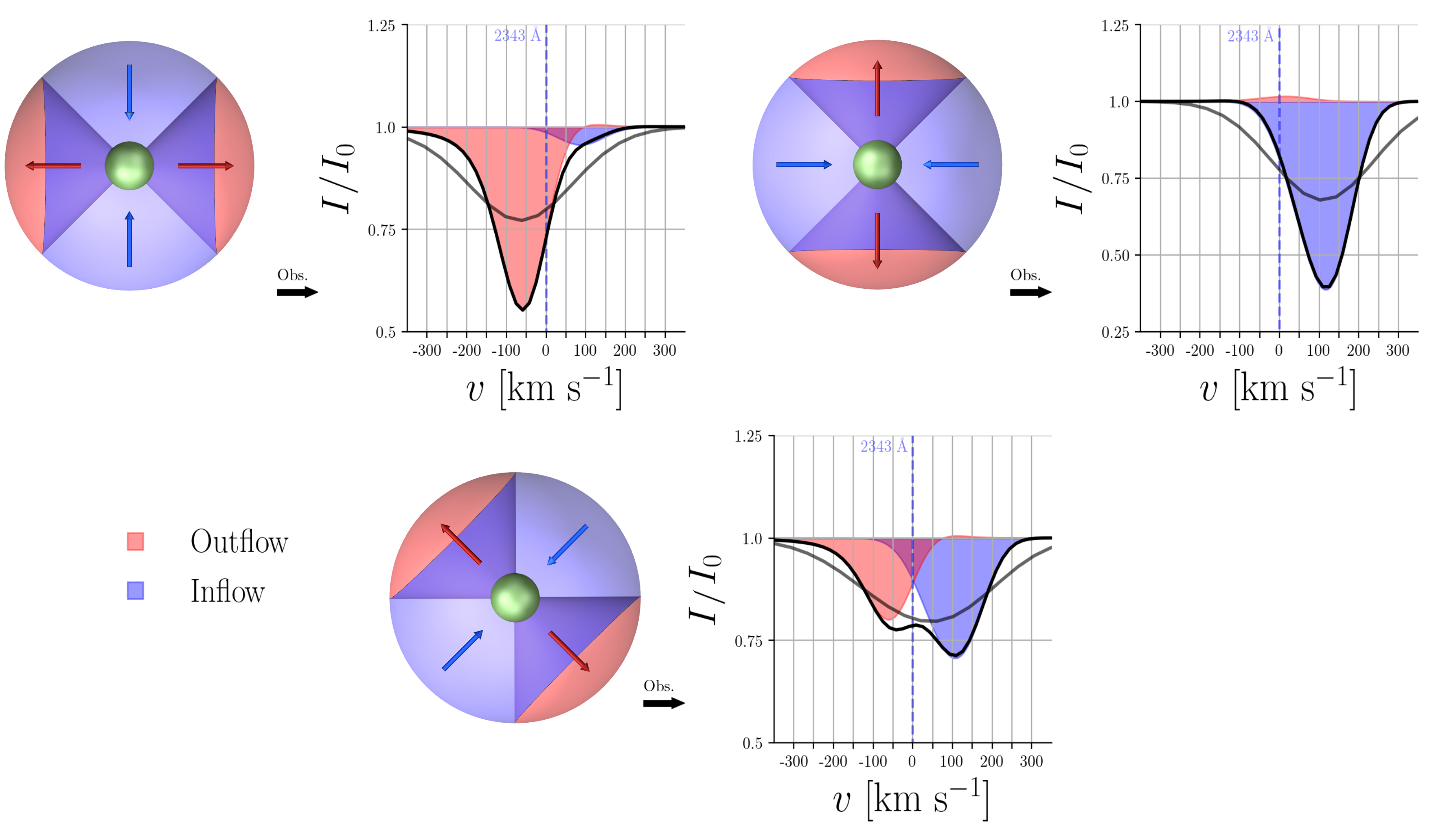}
 \caption{Orientation effects on galactic systems with inflows and outflows.  All diagrams are to be viewed from the right.  Inflowing gas is shown in light blue while outflowing gas is shown in salmon.  The would be spectrum for each flow is shaded in matching color.  The composite spectrum is shown at a resolution of $50 \ \rm{km} \ \rm{s}^{-1}$ (black) and $120 \ \rm{km} \ \rm{s}^{-1}$ (dark grey).  \textbf{\emph{Upper Left}} Looking down the line of site into an outflow with inflows moving perpendicular to the line of sight. \textbf{\emph{Upper Right}} Looking down the line of sight of an inflow with an outflow occurring perpendicular to the line of sight.  \textbf{\emph{Bottom Middle}} Both inflow and outflow are observed edge-on. The remaining parameters for these profiles are as follows.  Outflows: $\gamma = 1.0$, $\delta = 3.0$, $\log{(N \ [\rm{cm}^{-2}])}=15$, $v_0 = 25 \ \rm{km \ s^{-1}}$, $v_w = 500 \ \rm{km \ s^{-1}}$, $R_{AP} > R_W$, $\kappa = 0$, and $f_c = 1.0$.  Inflows: $\gamma = 0.5$, $\delta = 1.5$, $\log{(N \ [\rm{cm}^{-2}])}=15$, $v_0 = 15 \ \rm{km \ s^{-1}}$, $v_w = 200 \ \rm{km \ s^{-1}}$, $R_{AP} > R_W$, $\kappa = 0$, and $f_c = 1.0$.}
   \label{f7}
\end{figure*} 

Realistically, both accretion and feedback occur simultaneously in galaxies - for example, during gas recycling when material expelled by the galaxy is returned via gravity.  This is a fair assumption given the high detection rate of outflows and the necessity for inflows brought on by simulations.  In fact, it has been suggested that the presence of outflows can mask inflows in spectral lines preventing the latter's detection (e.g., \citealt{Martin2012}).  This can foreseeably occur if the symmetric, with respect to line center, emission profile of an outflow fills in the weaker absorption feature of an inflow.  Since numerical simulations show that outflows typically emerge in the direction perpendicular to the face of the disc while inflows move in the plane of the disc, orientation should also play a factor.

In this section, we attempt to explore which conditions favor the detection of inflows in galactic fountains (i.e., galactic systems with both inflows and outflows).  To prevent the masking scenario mentioned above, we have chosen to model the Fe II 2343\AA\ line.  This line has two strong fluorescent channels at wavelengths 2365\AA\ and 2381\AA, and will therefore incur very little resonant emission (see \citealt{Scarlata2015} for a detailed description of the role fluorescence plays in resonant scattering).  In addition, ignoring the emission component will also allow us to much more easily model the simultaneous outflowing and inflowing of a single ionic species in SALT.  In fact, under these conditions we can compute the spectrum from each model (inflow and outflow) independently and take the final spectrum as the superposition; otherwise, more modeling would be necessary to account for photons which are scattered by both the outflow and inflow.  We also note that this situation can be enforced observationally by limiting the portion of the outflow captured by the observing aperture.  By limiting the field of view to exclude the outflow except along the line of sight, one can effectively eliminate the entire emission component of a line profile (see \citealt{Scarlata2015} for justification).  This technique is feasible with IFU spectroscopy where one can obtain spectra from sub images.     

We explore how the line profile of a galactic fountain changes with orientation in Figure~\ref{f7}.  We show both a medium ($50 \ \rm{km} \ \rm{s}^{-1}$) and a low ($120 \ \rm{km} \ \rm{sec}^{-1}$) resolution spectra which has been smoothed with a Gaussian kernel.  The latter is about the high resolution limit of the study by \cite{Martin2012}.  In this setup, we are assuming that the outflows and inflows occupy separate regions of the CGM with inflows moving perpendicular to a biconical outflow.  Specifically, the models are made by generating the spectrum of a spherical inflow, removing the contribution of a biconical inflow from that spectrum, and then replacing it with a spectrum produced by a biconical outflow refilling that space.  The remaining volume filled by inflowing gas is roughly the shape of a torus.  Leaving such a large volume for the inflows represents a best case scenario for observing them.   As expected, the orientation (compare the top two profiles) strongly affects our ability to detect an inflow at all resolutions: If there is no gas moving along the line of sight then we cannot detect any redshifted absorption, and thus have no possible way to identify an inflow given the weak emission component for this line.  In the bottom panel, the low resolution spectrum of both the outflow and inflow appears to have merged together making it impossible to distinguish the inflow from another absorption feature such as absorption occurring in the ISM.  This appears to be another form of masking by the outflow since this would not be the case if the outflow were absent (see the blue profile).  There does appear to be hope of identifying the inflow at $50 \ \rm{km} \ \rm{s}^{-1}$ resolution, however, since the asymmetric redshifted absorption feature becomes visible.

\begin{figure*}[]
 \centering
\includegraphics[scale=0.83]{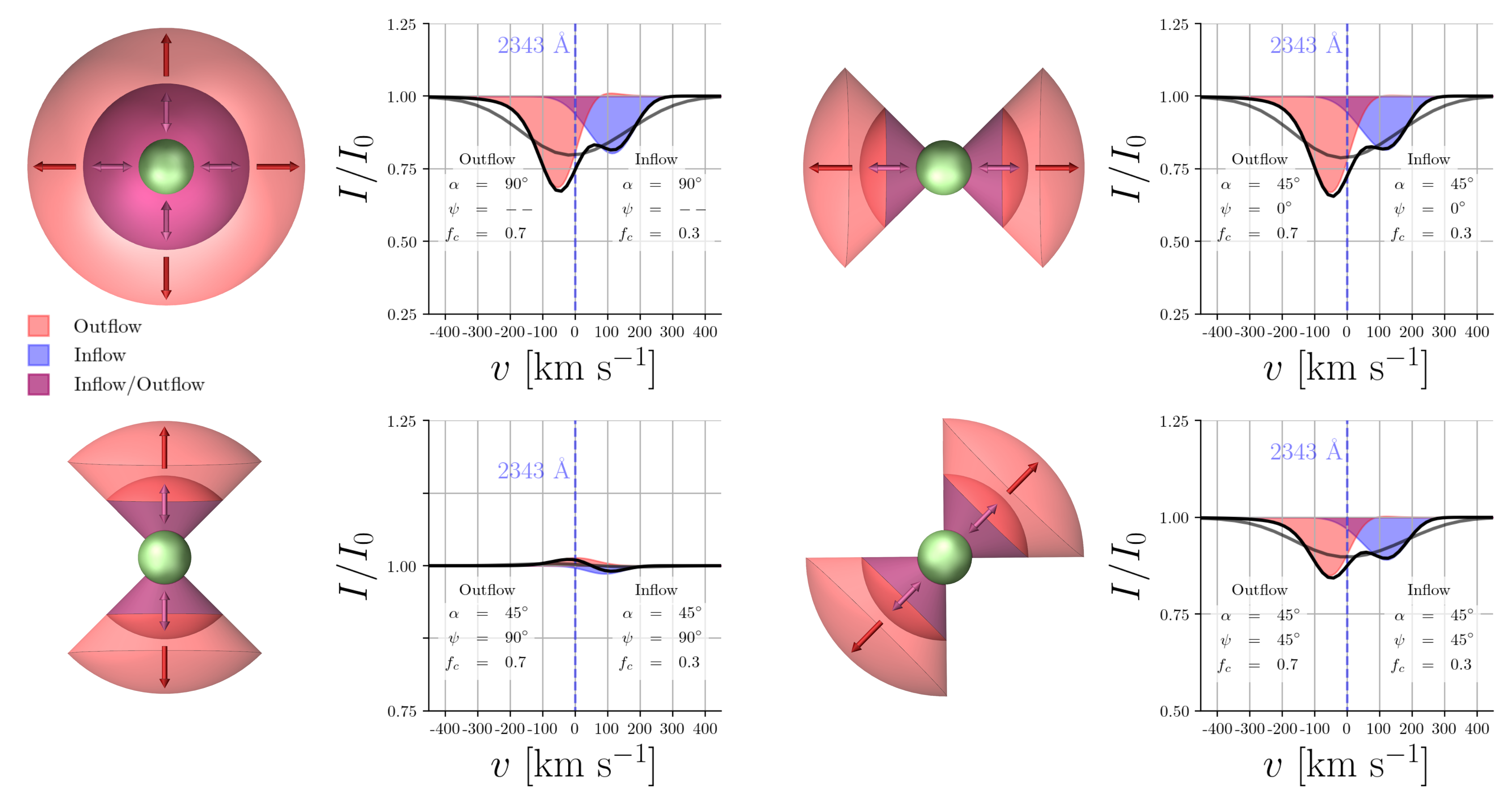}
 \caption{Galactic fountains with inflows represented by streams observed in the Fe II 2343\AA \ line viewed at different orientations and geometries.  The Fe II 2343\AA \ line has a strong fluorescent component (not shown) and therefore has reduced resonant emission.  All diagrams are to be viewed from the right.  Inflowing gas is shown in light blue, outflowing gas is shown in salmon, and regions of overlap are shown in violet.  The resulting spectra for each type of flow is shaded in the matching color.  The composite spectrum is shown at a resolution of $50 \ \rm{km} \ \rm{s}^{-1}$ (black) and $120 \ \rm{km} \ \rm{s}^{-1}$ (dark grey).  \textbf{\emph{Top Left}} A spherical fountain.  \textbf{\emph{Top Right}} A biconical fountain observed parallel to the line of sight.  \textbf{\emph{Bottom Left}} A biconical fountain oriented perpendicular to the line of sight. \textbf{\emph{Bottom Right}} A biconical fountain observed edge on.  The remaining parameters are as follows.  Outflows: $v_0 = 25 \ [\rm{km} \ \rm{s}^{-1}]$, $v_{\infty} = 500 \ [\rm{km} \ \rm{s}^{-1}]$, $R_{AP} >  R_W$, $\log{(N \ [\rm{cm}^{-2}])}=15$, $\kappa = 0$, $\delta = 3.0$, and $\gamma = 1.0$.  Inflows: $v_0 = 15 \ [\rm{km} \ \rm{s}^{-1}]$, $v_{\infty} = 200 \ [\rm{km} \ \rm{s}^{-1}]$, $R_{AP} >  R_W$, $\log{(N \ [\rm{cm}^{-2}])}=15$, $\kappa = 0$, $\delta = 1.5$, and $\gamma = 0.5$.      }
   \label{f8}
\end{figure*}

In the case of a true fountain, where warm ($5050\rm{K} < T < 2\times 10^4\rm{K}$) ejected gas doesn't have enough energy to escape the gravitational potential well, large regions of the CGM can have both an inflowing and outflowing component (see \citealt{Kim2018}).  We attempt to model such situations in Figure~\ref{f8}, where we show various outflows with an inflowing gas component occupying the same general volume.  For each scenario, we have chosen covering fractions for the outflows, $f_{c,o}$, and inflows, $f_{c,i}$, such that $f_{c,o} + f_{c,i} = 1$ and $f_{c,i} <f_{c,o}$ with the order representing the fact that inflows are typically expected to have smaller covering fractions than outflows \citep{Kimm2011}.  Once again, we are not able to identify inflows at the $120 \ \rm{km} \ \rm{sec}^{-1}$ resolution level given our model constraints, but are able to at the $50 \ \rm{km} \ \rm{s}^{-1}$ level.  This strongly suggests that the number of inflows detected in large low resolution surveys is undercounted, and can be improved by moving to higher, but still obtainable resolutions by active instruments. 

\section{Discussion and Conclusions}

While the detection rate for accretion or galactic inflows is much lower than that of outflows in absorption line studies (see \citealt{Martin2012} and \citealt{Rubin2014}), the important role inflows play in models of galaxy formation suggests that they should be just as common as outflows.  This fact implies that the effects of inflows on spectra lines are either weak (i.e., inflows have low covering fractions and/or low column densities) or hidden (i.e., the effects of inflows are masked by outflows and/or are mistaken for the ISM).  

In regards to the latter, given the available constraints on inflows, our models suggest that we should be able to resolve the properties of inflows with observations reaching spectral resolution of $\approx 50\ \rm{km} \ \rm{sec}^{-1}$ or $R=6000$. In addition to high spectral resolution, the detection of outflows/inflows requires spectra with high signal-to-noise ratio in the continuum\footnote{We do not explore the dependency on signal-to-noise here, but extensive discussion can be found in the literature.}.  The COS spectrograph on board of the Hubble Space Telescope provides access to the required spectral resolution, but has no multiplexing capabilities, and is limited in the fluxes that it's able to observe.  Observing large samples of objects at the required resolution and SNR is thus a challenging task.  Future space-based missions, such as the large infrared/optical/ultraviolet telescope recommended by the Astro2020 Decadal Survey, will substantially improve the situation for low-redshift galaxies.  For galaxies at $z>0.6$ some of the resonant spectral features  are redshifted in the optical and can be observed from the ground.  A number of spectrographs on large-aperture telescopes provide both the resolution and the mutiplexing capability that allow the study of statistical samples of galaxies in reasonable time frames.  Examples include FLAMES, the  multi-object, intermediate and high resolution spectrograph of the Very Large Telescope, and the IMACS multi-object spectrograph on the 6.5m Magellan telescope.  Additionally, focusing on ionic transition lines with strong fluorescent channels which naturally have less resonant emission (such as the Fe~II~2343\AA \ transition studied in this work) will mitigate the effect outflows have on masking inflows.  

The SALT model presented in this work will be invaluable in constraining the properties of gas inflows and outflows identified in high spectral-resolution data.

\acknowledgments
\section*{Acknowledgements}
The authors would like to thank Jacob Miller for graciously helping to create the figures in this work.  The authors thank the anonymous referee for an in depth reading of the manuscript and for providing valuable insight for parameter constraints.

\appendix
In Table~\ref{tab:pars} we report the primary SALT parameters with their symbol and their definition. The second half of the table defines additional support parameters used in the calculation of  models in Section~2. 

\begin{table*}[t]
	\center
	\caption{List of SALT parameters and definitions. }
	\begin{tabular}{ccc}\hline\hline
		Symbol& \ \ \ \ \ \ \ \ \ \ \ \ \ \ \ \ \ \ \ \ \ \ \ \ \ \ \  \ \ \ \ \ \ \ \ \ \ \ \ \ \ \ \ \ \ \ \ \ \ \ \ \ \ \   \ \ \ \ \ \ \ \ \ \ \ \ \ \ \ \ \ \ \ \ \ \ \ \ \ \ \  Definition\\
		\hline
		&Outflow&Inflow\\
		\hline 
		$R_{SF}$ & Radius of continuum source& $\cdot\cdot$\\
		$R_W$ & Wind radius&$\cdot\cdot$\\
		$R_{AP}$ & Aperture radius&$\cdot\cdot$\\
		$\alpha$&Half opening angle of the biconical flow&$\cdot\cdot$  \\ 
		$\psi$ &Orientation angle of biconical flow w.r.t. the l.o.s. &$\cdot\cdot$\\  
		$v_0$ &Launch velocity of outflow & Scales inflowing velocity field \\
		$v_{\infty}$ &Terminal velocity of flow&$\cdot\cdot$  \\
		$v_{ap}$ &Value of velocity field at $R_{AP}$&$\cdot\cdot$ \\
		$\gamma$ &Velocity field power law index &$\cdot\cdot$\\
		$\tau_0$ &Optical depth of flow&$\cdot\cdot$ \\  
		$\tau$ & $\tau_0$ divided by $f_{ul} \lambda_{ul}$&$\cdot\cdot$ \\
		$\delta$ &Density field power law index &$\cdot\cdot$\\
		$\kappa$ &Dust opacity multiplied by $R_{SF} n_{\rm{dust},0}$&$\cdot\cdot$ \\
		$f_c$ &Covering fraction inside the biconical flow&$\cdot\cdot$ \\
		$n_0$ &Number density at $R_{SF}$&$\cdot\cdot$ \\
		\hline
		$x$/$y$/$y_\infty$& observed velocity/shell velocity/terminal velocity normalized by $v_0$&$\cdot\cdot$\\
		$\phi$ & Angle between the velocity and incoming photon ray&$\cdot\cdot$\\
	$\Omega_x$&Surface of constant observed velocity $x$ &$\cdot\cdot$\\
	$\Gamma_x$ & Intersection of $\Omega_x$ with the $s\xi$-plane&$\cdot\cdot$  \\
	$\Xi$& $\xi$ coordinate of $\Gamma_x$&$\cdot\cdot$ \\
	$S$& $s$ coordinate of $\Gamma_x$&$\cdot\cdot$ \\
		$\tau_S$&Sobolev optical depth&$\cdot\cdot$\\
		$\bar{\tau}_S$&NA&$\tau_S$ defined for more than one Sobolev point\\
		$f_{lu}$&Oscillator strength for the $lu$ transition&$\cdot\cdot$\\
		$\lambda_{lu}$&Wavelength for the $lu$ transition&$\cdot\cdot$\\
			$\Theta_C$& Max angle for continuum absorpiton&$\cdot\cdot$\\
			$\Theta_{\rm{red}}$&Occultation scale factor for red emission&NA\\
			$\Theta_{\rm{blue}}$&NA&Occultation scale factor for blue emission\\
			$f_g$ & Geometric scale factor for a biconical wind geometry&$\cdot\cdot$\\
		\hline 
		\vspace{0.4cm}
	\end{tabular}
	\label{tab:pars}
\end{table*}  

\begin{table*}[t]
\caption{Atomic Data for Fe II ion.  Data taken from the NIST Atomic Spectra Database$^{\MakeLowercase{a}}$.}
\begin{tabular}{P{.7cm}|P{2.3cm}|P{1.4cm}|P{1.6cm}| P{2.8cm}|P{.9cm}|P{2.6cm}|P{2.8cm}}\hline\hline
Ion & Vac. Wavelength&  $A_{ul}$&$f_{ul}$& $E_{l}-E_{u}$&$g_l-g_u$&Lower Level&Upper Level\\
&\AA& $s^{-1}$ &&$eV$&&Conf.,Term,J&Conf.,Term,J\\[.5 ex]
\hline 
 Fe II& $2343.49$ &$1.73\times 10^{8}$&$1.14\times 10^{-1}$&$0.0000-42658.2444 $&$10-8$ &$3d^6({}^3H)4p, a{}^6D, 9/2$&$3d^6({}^5D)4p, z{}^6P^o, 7/2$ \\
&$2364.83$ & $5.9\times 10^{7}$ &$4.95\times 10^{-2}$&$384.7872-42658.2444 $&$8-8$&$3d^6({}^5D)4s, a{}^6D, 7/2$&$3d^6({}^5D)4p, z{}^6P^o, 7/2$  \\
& $2380.76$ & $3.10\times 10^{7}$ &$3.51\times 10^{-2}$&$667.6829-42658.2444$&$6-8$&$3d^6({}^5D)4s, a{}^6D, 5/2$&$3d^6({}^5D)4p, z{}^6P^o, 7/2$ \\
& $2382.04$ & $3.13\times 10^{8}$ &$3.2\times 10^{-1}$&$0.0000-41968.0698$&$10-12$&$3d^6({}^5D)4s, a{}^6D, 9/2$&$3d^6({}^5D)4p, z{}^6F^o, 11/2$ \\[1ex]
\hline
\end{tabular}
\label{tab:atomicdata}
\\$^{\text{a}}$ http://www.nist.gov/pml/data/asd.cfm
\end{table*}

\newpage

\bibliographystyle{jphysiscsB} 
\bibliographystyle{apj} 
\bibliography{doc}

\end{document}